\documentclass{mn2e}
\usepackage{txfonts}
\usepackage{graphicx}
\usepackage{amssymb}
\usepackage{color}

\newcommand{\be}{\begin{equation}}
\newcommand{\ee}{\end{equation}}
\newcommand{\etal}{et al.}
\newcommand{\rxjw}{RX~J1856.5$-$3754}
\newcommand{\xmm}{\textit{XMM-Newton}}
\newcommand{\zg}{z_g}
\newcommand{\Teff}{T_{\rm eff}}
\newcommand{\rem}{R^{\rm em}}
\newcommand{\tinfty}{T^{\infty}}
\newcommand{\rinfty}{R^{\infty}}

\newcommand{\rinftyo}{R^{\infty}_{\rm opt}}
\newcommand{\vecB}{\mathbf B}
\newcommand{\ThetaB}{\Theta_B}
\newcommand{\Omegas}{\mathbf \Omega_{\rm s}}
\newcommand{\thetap}{\alpha}
\newcommand{\phip}{\phi_{\rm p}}
\newcommand{\thetar}{\zeta}
\newcommand{\veck}{\mathbf k}
\newcommand{\thetak}{\theta_k}
\newcommand{\phik}{\phi_k}
\newcommand{\muk}{\mu_k}
\newcommand{\vecm}{\mathbf m}
\newcommand{\thetam}{\theta_m}
\newcommand{\phim}{\phi_m}
\newcommand{\vecx}{\mathbf x}
\newcommand{\hatx}{\hat{\mathbf x}}
\newcommand{\hatz}{\hat{\mathbf z}}
\newcommand{\vecxn}{\mathbf x_{\rm n}}
\newcommand{\hatxn}{\hat{\mathbf x}_{\rm n}}
\newcommand{\hatzn}{\hat{\mathbf z}_{\rm n}}
\newcommand{\chisq}{\chi^2}

\title[Constraining the geometry of RX~J1856.5$-$3754]{
Constraining the geometry of the neutron star RX~J1856.5$-$3754}
\author[W.C.G. Ho]{Wynn C. G. Ho$^1$\thanks{email: wynnho@slac.stanford.edu} \\
$^1$Harvard-Smithsonian Center for Astrophysics, 60 Garden St., Cambridge,
MA, 02138, USA}

\begin{document}
\pagerange{\pageref{firstpage}--\pageref{lastpage}} \pubyear{2007}
 
\maketitle

\label{firstpage}

%%%%%%%%%%%%%%%%%%%%%%%%%%%%%%%%%%%%%%%%%%%%%%%%%%%%%%%%%
\begin{abstract}
RX~J1856.5$-$3754 is one of the brightest, nearby isolated neutron stars,
and considerable observational resources have been devoted to its study.
In previous work, we found that our latest models of a magnetic, hydrogen
atmosphere matches well the entire spectrum, from X-rays to optical
(with best-fitting neutron star radius $R\approx 14$~km, gravitational redshift
$\zg\sim 0.2$, and magnetic field $B\approx  4\times 10^{12}$~G).
A remaining puzzle is the non-detection of rotational modulation of the X-ray
emission, despite extensive searches.
The situation changed recently with \xmm\ observations that uncovered 7~s
pulsations at the $\approx 1\%$ level.
By comparing the predictions of our model (which includes simple dipolar-like
surface distributions of magnetic field and temperature) with the observed
brightness variations, we are able to constrain the geometry of
RX~J1856.5$-$3754, with one angle $< 6^\circ$ and the other angle
$\approx 20-45^\circ$,
though the solutions are not definitive given the observational and model
uncertainties.
These angles indicate a close alignment between the rotation and
magnetic axes or between the rotation axis and the observer.
We discuss our results in the context of RX~J1856.5$-$3754 being a normal
radio pulsar and a candidate for observation by future X-ray polarization
missions such as {\it Constellation-X} or {\it XEUS}.
\end{abstract}


%%%%%%%%%%%%%%%%%%%%%%%%%%%%%%%%%%%%%%%%%%%%%%%%%%%%%%%%%%%%%%%%%%%
\section{Introduction} \label{sec:intro}

Seven candidate isolated, cooling neutron stars (INSs) have
been identified by the ROSAT All-Sky Survey
(see Kaspi, Roberts, \& Harding~2006; Haberl~2007;
van Kerkwijk \& Kaplan~2007, for recent reviews).
These objects share the following properties: (1) high
X-ray to optical flux ratios of $\log(f_{\rm X}/f_{\rm optical})\sim 4-5.5$,
(2) soft X-ray spectra that are well described by blackbodies with
$kT\sim 50-100$~eV, (3) relatively steady X-ray flux over long timescales,
and (4) lack of radio pulsations.

For the particular INS \rxjw, X-ray and optical/UV data can be well-fit
by two corresponding blackbody spectra: the X-ray spectrum by a blackbody
with $k\tinfty_{\rm X}=63$~eV and emission size
$\rinfty_{\rm X}=5.1\,(d/\mbox{140 pc})$~km,
and the optical/UV spectrum by a blackbody with $k\tinfty_{\rm opt}=26$~eV,
and $\rinfty_{\rm opt}=21.2\,(d/\mbox{140 pc})$~km
(Burwitz \etal~2001; van Kerkwijk \& Kulkarni~2001a; Drake \etal~2002;
see also Pavlov, Zavlin, \& Sanwal~2002; Tr\"{u}mper \etal~2004; Walter~2004),
where $\tinfty=\Teff/(1+\zg)$, $\rinfty=\rem(1+\zg)$,
$\rem$ is the physical size of the emission region, and $d$ is the distance.
The gravitational redshift $\zg$ is given by $(1+\zg)=(1-2GM/Rc^2)^{-1/2}$,
where $M$ and $R$ are the mass and radius of the NS, respectively.
The high temperature, small area X-ray blackbody suggests
a small hot spot on the NS surface, while the optical blackbody with
$\rinftyo$ is the remaining large, cool surface.

Even though blackbody spectra fit the data, one expects NSs to possess
atmospheres of either heavy elements (due to debris from the progenitor)
or light elements (due to gravitational settling or accretion).
The lack of any significant spectral features in
the X-ray spectrum argues against a heavy element atmosphere (Burwitz
\etal~2001, 2003), whereas non-magnetic or fully-ionized magnetic
hydrogen atmosphere spectra do not provide a good fit
(Pavlov \etal~1996; Pons \etal~2002; Burwitz \etal~2001, 2003).
For the last case, it is important to note that,
since $kT\sim\mbox{tens of}$~eV for
\rxjw\ and the ionization energy of hydrogen at $B=10^{12}$~G is
160~eV, the presence of neutral atoms must be accounted for in the
magnetic hydrogen atmosphere models; the opacities are sufficiently
different from the fully ionized opacities that they can change the
atmosphere structure and continuum flux (Ho \etal~2003; Potekhin \etal~2004),
which can affect fitting of the observed spectra (see, e.g., Ho \etal~2007).

Another issue that may argue against the two-temperature blackbody model is
the the non-detection, until recently, of X-ray pulsations due to the rotation
of the NS (down to the 1.3\% level;
Drake \etal~2002; Ransom, Gaensler, \& Slane~2002; Burwitz \etal~2003;
Zavlin~2007).
As discussed in the important work by Braje \& Romani~(2002), very low
rotational pulsations could be due to a very close alignment between the
rotation and magnetic axes and/or an unfavorable viewing geometry.
Assuming blackbody emission, they find this is possible in
$\sim 5\%$ of viewing geometries (with $R=14$~km).
However, more realistic atmosphere emission can lead to enhancement of
pulsations, as well as an energy-dependence of the pulsations (see below).

Thus there existed three major observational and theoretical inconsistencies:
(1) the inferred emission size from blackbody fits are either much smaller
($\rinfty\sim 5$~km from the X-ray data) or much larger ($\rinfty\sim 20$~km
from the optical/UV data) than the canonical NS radius of $10-12$~km, even
after correcting for the redshift,
(2) blackbodies fit the spectrum much better than realistic atmosphere models,
and (3) strong upper limits on X-ray pulsations suggest \rxjw\ may have
a largely uniform temperature over the entire NS surface.
In response to these problems, we applied our latest magnetic,
partially-ionized hydrogen atmosphere models and obtained a good fit to the
entire multi-wavelength spectrum of \rxjw\ (Ho \etal~2007; hereafter H07).
Besides using the more realistic atmosphere, the fit requires a smaller
emission size $\rem\approx 14$~km (as compared to the blackbody
models); if interpreted as emission from the entire NS, the size can be
satisfied by a stiff but standard equation of state
(see, e.g., Lattimer \& Prakash~2007).  Furthermore, we found geometries
that produce pulsations near the observed limits.
Because of observational uncertainties and the computationally tedious
task of constructing a complete grid of models, the results may not be
unique.  Nevertheless, our model represents the most self-consistent
picture for explaining all the observations of \rxjw.

Very recently, \xmm\ observations uncovered pulsations from \rxjw\
with a period of 7~s (pulse amplitude $\approx 1\%$) and an upper limit
on the period derivative; by
assuming vacuum magnetic dipole braking, this implies $B<10^{14}$~G
(Tiengo \& Mereghetti~2007; hereafter TM07).
The detection of pulsations allows us to constrain the geometry
(i.e., two angles $\alpha$ and $\zeta$, where $\alpha$ is the angle between
the rotation and magnetic axes and $\zeta$ is the angle between the
rotation axis and the direction to the observer)
through a comparison of the observed brightness variations
($\equiv$~light curves) with those implied by the model of H07;
the latter are obtained by computing the (rotation) phase-dependent
spectra from the entire NS surface.

Spectra from the whole NS surface are necessarily model-dependent
(see, e.g., Zavlin \etal~1995; Zane \etal~2001; Ho \& Lai~2004;
Zane \& Turolla~2006), as the magnetic field and
temperature distributions over the NS surface are unknown.
Magnetic field variations over the surface will induce surface
temperature variations (Greenstein \& Hartke~1983).
Furthermore, radiation from the surface of a magnetic NS with an
atmosphere differs significantly from that implied by isotropic blackbody
emission: the emission depends on the direction of the local surface
magnetic field (Shibanov \etal~1992; Pavlov \etal~1994).
Indeed, there have been recent works attempting to fit magnetic atmosphere
spectra to observations of NSs other than \rxjw.
Lloyd \etal~(2003) fit the spectrum and light curve of PSR~0656+14,
while Zavlin \& Pavlov~(2004), using a different model, do the same for
PSR~B0950+08.
Zane \& Turolla~(2006), using the method of Lloyd \etal~(2003)
to produce an extensive library of light curves for fully ionized
hydrogen atmospheres, fit the light curves of several INSs.

An outline of the paper is as follows.  In Section~\ref{sec:model},
we describe the model used to produce light curves of \rxjw.
Light curves for arbitrary angles ($\alpha,\zeta$) are discussed in
Section~\ref{sec:results}.
The observations and fits to the observations are shown in
Section~\ref{sec:rxj1856}.
We summarize and discuss our results in Section~\ref{sec:discussion}.

%%%%%%%%%%%%%%%%%%%%%%%%%%%%%%%%%%%%%%%%%%%%%%%%%%%%%%%%%%%%%%%%%%%
\section{Model for Neutron Star Emission} \label{sec:model}

We adopt a relatively simple model (see also Zavlin \etal~1995;
Lloyd \etal~2003) for the
surface magnetic field $B$ and effective temperature $\Teff$ distributions:
we assume the surface is symmetric (in $B$ and $\Teff$) about the magnetic
equator and divide the hemisphere into four magnetic colatitudinal regions.
We generate (local) atmosphere models (see H07, and references
therein, for details;
see Zavlin \& Pavlov~2002, for a review of atmosphere modeling)
for each region with the parameters given in
Table~\ref{tab:nssurf}, where $\ThetaB$ is the angle between the local
magnetic field and the surface normal.
Emission from any point within a colatitudinal region is given by the
atmosphere model for that region.
Note that the magnetic field distribution is roughly dipolar.
The phase-resolved spectra and light curves from the entire NS surface is
then calculated by the method described below (see the analogous formalism
in Pavlov \& Zavlin~2000).
We assume $M=1.4 M_\odot$ and $R=14$~km (see H07).  This implies
$\zg=0.2$, and all energies quoted in this work have been redshifted
from the local NS frame by this factor.

%-------------------------------------------
\begin{table}
\caption{Parameters for Neutron Star Surface \label{tab:nssurf}}
\begin{tabular}{c c c c}
\hline
magnetic colatitude & $B$ & $\ThetaB$ & $\Teff$ \\
 (deg) & ($10^{12}$~G) & (deg) & ($10^5$~K) \\
\hline
0$-$10 & 6 & 0 & 7 \\
10$-$40 & 5 & 30 & 6 \\
40$-$70 & 4 & 60 & 5 \\
70$-$90 & 3 & 90 & 4 \\
\hline
\end{tabular}
\end{table}
%-------------------------------------------

\begin{figure}
\resizebox{\hsize}{!}{\includegraphics{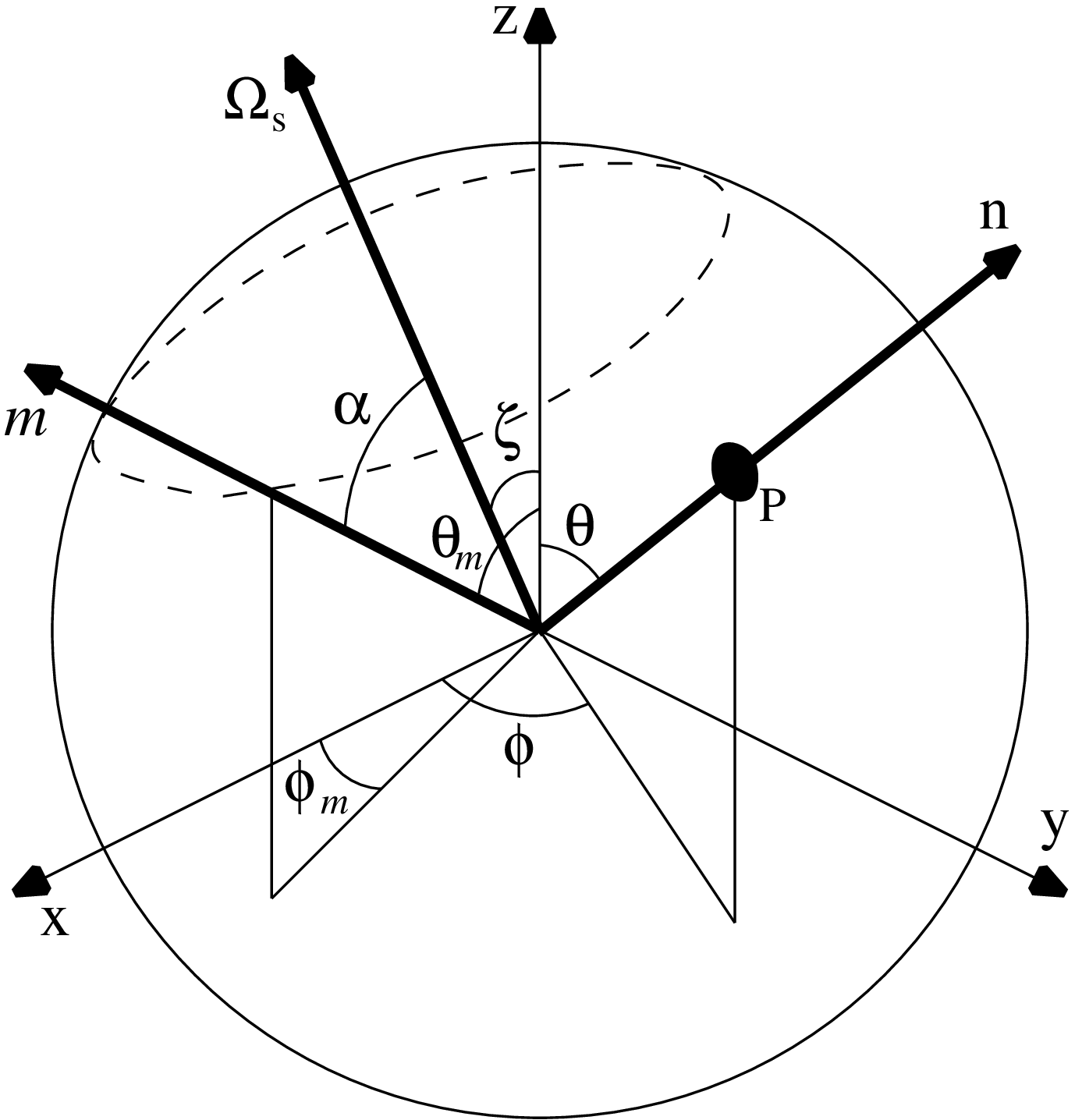}\includegraphics[width=0.6\textwidth]{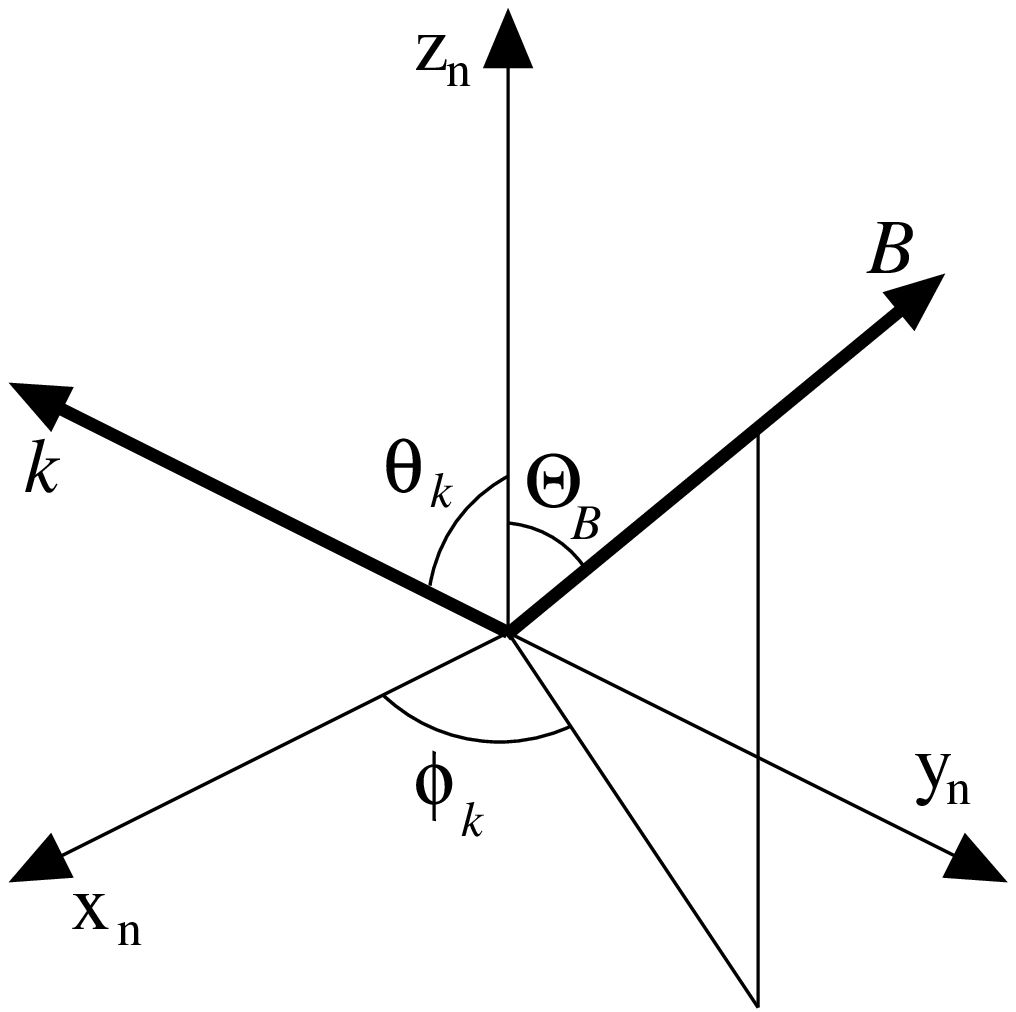}}
\caption{Coordinate axes and angles used to describe the pulsed emission.
The observer is in the $\hatz$-direction.
The dashed line indicates the position of the magnetic pole $\vecm$
as the neutron star rotates about $\Omegas$.
}
\label{fig:angles}
\end{figure}

The observed emission from a local region (e.g., the hot magnetic polar cap)
of a rotating NS depends on two angles: the angle $\alpha$ between the
rotation and magnetic pole axes and the angle $\zeta$ between the
rotation axis and the line-of-sight to the observer.
In the coordinate frame $\vecx$, where $\hatz$ is the direction to the
observer and the spin axis $\Omegas$ is in the $\hatx\hatz$-plane,
the polar angle $\thetam$ and azimuthal angle $\phim$ of the magnetic
pole $\vecm$ are given by
\begin{eqnarray}
\cos\thetam & = & \cos\phip\sin\thetap\sin\thetar + \cos\thetap\cos\thetar \\
\tan\phim & = & \frac{\sin\phip\sin\thetap}{
 \cos\phip\sin\thetap\cos\thetar - \cos\thetap\sin\thetar},
\end{eqnarray}
respectively (see Figure~\ref{fig:angles}).
$\phip$ ($=\Omega_{\rm s} t$) is the rotation phase.
Note that the light curves are symmetric under the interchange of
$\alpha$ and $\zeta$, since the brightness depends only on the angle
$\thetam$ between the magnetic and line-of-sight axes.
At the emission-point P [=P($\theta,\phi$)] on the NS surface
(see Figure~\ref{fig:angles}), the direction of the local magnetic
field $\vecB$ (assuming a magnetic dipole geometry) is given by
\be
\cos\ThetaB = \cos(\phim-\phi)\sin\theta\sin\thetam +\cos\theta\cos\thetam
\ee
\be
\tan\phik = \frac{\sin(\phim-\phi)\sin\thetam}{
\cos(\phim-\phi)\cos\theta\sin\thetam-\sin\theta\cos\thetam}.
\ee
$\phik$ is the azimuthal angle between $\vecB$ and the photon wavevector
$\veck$ and is defined in the coordinate frame $\vecxn$, where $\hatzn$
is the direction of the surface normal and $\veck$ is in the
$\hatxn\hatzn$-plane.

%-------------------------------------------
%-------------------------------------------

To determine the flux spectrum $F_E$ for a given rotation phase,
we first compile a table of specific intensities
$I_E(\thetak,\phik)$,
where $\thetak$ is the angle between $\veck$ and $\hatzn$.
$I_E$ are computed from atmosphere models described in H07;
these models depend on the local effective
temperature and magnetic field direction $\ThetaB$ and strength
(see Table~\ref{tab:nssurf}).
The flux spectrum can then be calculated from
\be
F_E = \left(\frac{\rinfty}{d}\right)^2\frac{1}{(1+\zg)^3} \int_0^{2\pi}d\phik
\int_0^1d\muk \,\muk I_E(\thetak,\phik),
\ee
where $\muk=\cos\thetak$.
In the absence of the bending of the path of light due to gravity, the
polar angles of the emission-point and the photon wavevector are the same,
i.e., $\theta=\thetak$.
Gravitational light-bending causes more of the NS surface to be visible
(for $M=1.4M_\odot$ and $R=14$~km, $\theta\le 115^\circ$ as compared to
$\le 90^\circ$ without light-bending).
We use the approximate relation between $\theta$ and $\thetak$ given
in Beloborodov~(2002)
\be
\cos\theta = \frac{\cos\thetak - 2GM/c^2R}{1 - 2GM/c^2R},
\ee
which deviates from the exact relation given in Pechenick, Ftaclas,
\& Cohen~(1983) by $\lesssim 1\%$ for our chosen $M$ and $R$.
The light curves are then computed by integrating the photon
count spectra over the energy range of the \xmm\ observations (see TM07):
0.15$-$1.2~keV for the total X-ray spectrum, 0.15$-$0.26~keV for the
soft band, and 0.26$-$1.2~keV for the hard band.

%%%%%%%%%%%%%%%%%%%%%%%%%%%%%%%%%%%%%%%%%%%%%%%%%%%%%%%%%%%%%%%%%%%
\section{Results} \label{sec:results}

\begin{figure}
\resizebox{\hsize}{!}{\includegraphics{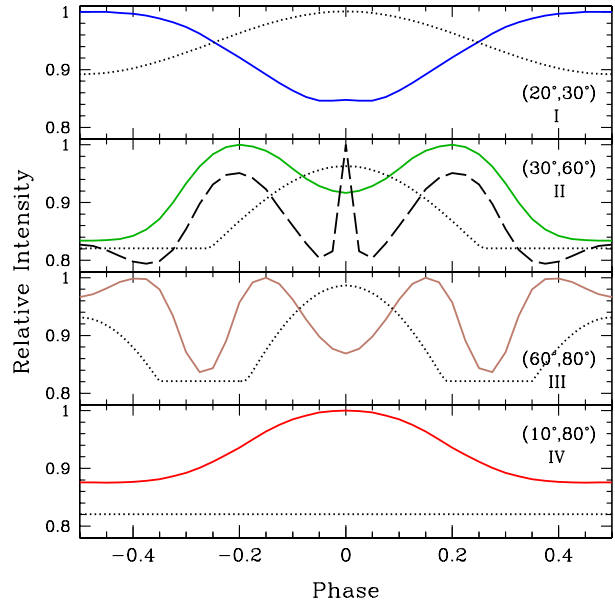}}
\caption{Light curves for different geometries ($\alpha,\zeta$):
class~I with (20$^\circ$,30$^\circ$), class~II with (30$^\circ$,60$^\circ$),
class~III with (60$^\circ$,80$^\circ$), and class~IV with
(10$^\circ$,80$^\circ$).
$\alpha$ is the angle between the spin and magnetic axes, and
$\zeta$ is the angle between the spin axis and the line-of-sight.
The four classes are defined in Beloborodov~(2002).
The solid lines are for the magnetic model described in the text
[dashed line is for (50$^\circ$,50$^\circ$)],
while the dotted lines are analytic light curves (scaled arbitrarily
in amplitude) for isotropic emission from two antipodal hot spots
(see Beloborodov~2002).
}
\label{fig:pulse}
\end{figure}

Figure~\ref{fig:pulse} shows the light curves of our NS model
for various geometries ($\alpha,\zeta$).
We also plot the analytic light curves from Beloborodov~(2002) for
isotropic emission from two antipodal hot spots
(see Zavlin \& Pavlov~1998; Bogdanov, Rybicki, \& Grindlay~2006, for
examples of pulse profiles from non-magnetic hydrogen atmosphere hot spots).
We refer to the region at magnetic colatitude 0$-$10$^\circ$ as the
primary magnetic polar cap and the opposite pole as the secondary cap.
The classification scheme (for isotropically-emitting hot caps)
is defined in Beloborodov~(2002):
(I) only the primary cap is visible, and the pulse profile is purely
sinusoidal with a single peak,
(II) the secondary cap is seen around pulse minimum due to relativistic
light-bending, which reduces the strength of the modulation,
(III) the primary cap is not seen during a segment of the rotation,
and (IV) both spots are seen at all phases and thus there is no modulation.
Relativistic light-bending allows $\theta\leq 115^\circ$ to be visible;
thus, e.g., for (20$^\circ$,80$^\circ$), one pole will not be seen
during a portion of the rotation.

%-------------------------------------------
%-------------------------------------------

Several important features are evident from a comparison of magnetic
atmosphere emission to that of isotropic emission.
The ($\thetak,\phik$)-angular-dependence of the radiation (or beam pattern)
manifests as a narrow ``pencil-beam'' along the direction of the magnetic
field and a broad ``fan-beam'' at intermediate angles (see Pavlov \etal~1994;
Lloyd~2003, for beam patterns and spectra at various $\ThetaB$ and $\phik$).
As discussed in Pavlov \etal~(1994), the pencil-beam is the result of the
lower opacity at $\thetak\lesssim (E/E_B)^{1/2}$, where
$E_B=\hbar eB/m_{\rm e}c=11.6\,(B/10^{12}\mbox{ G})$~keV is the electron
cyclotron energy; the width of the pencil-beam is thus
$\sim (E/E_B)^{1/2}$, and the radiation is more strongly beamed at higher
magnetic fields.  For our case with $B\sim 4\times 10^{12}$~G, the width
is $\sim 6^\circ$.
This narrow beam is seen in the (50$^\circ$,50$^\circ$)-light curve plotted
in Figure~\ref{fig:pulse}, which is the only instance shown that has the
observer's line-of-sight exactly crossing the magnetic cap and coinciding
with the peak of the isotropic emission.
Also evident is the fan-beam (most obvious in the light curves of classes
II and III), which occur on either side of the magnetic cap and can increase
the number of light-curve peaks.
In addition, the anisotropic beam pattern (combined with the surface
temperature variation) can produce an apparent phase shift compared to
isotropic emission and modulation when an isotropic beam pattern
suggests none (c.f. class IV).

\begin{figure}
\resizebox{\hsize}{!}{\includegraphics{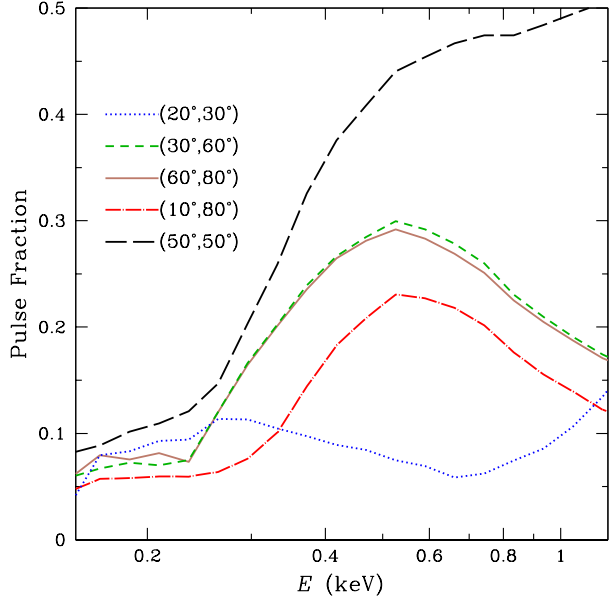}}
\caption{Pulse fractions as a function of energy for different geometries
($\alpha,\zeta$) = (20$^\circ$,30$^\circ$), (30$^\circ$,60$^\circ$),
(60$^\circ$,80$^\circ$), (10$^\circ$,80$^\circ$), (50$^\circ$,50$^\circ$).
$\alpha$ is the angle between the spin and magnetic axes, and
$\zeta$ is the angle between the spin axis and the line-of-sight.
}
\label{fig:pf}
\end{figure}

Figure~\ref{fig:pf} shows the pulse fractions,
$PF=(C_{\rm max}-C_{\rm min})/(C_{\rm max}+C_{\rm min})$, where $C$ is
the count spectrum, as a function of energy for various geometries.
The energy-dependence of the light curves and pulse fractions is due to
the energy-dependent beam patterns and the surface temperature and
magnetic field variations.
As noted in Zavlin \etal~(1995), the pulse fraction is lower at low
energies since the beam pattern is more isotropic.
Phase shifts between the peak of the low and high-energy light curves
can also occur.

%-------------------------------------------
%-------------------------------------------

%%%%%%%%%%%%%%%%%%%%%%%%%%%%%%%%%%%%%%%%%%%%%%%%%%%%%%%%%%%%%%%%%%%
\section{Comparison to \rxjw} \label{sec:rxj1856}

\begin{figure}
\resizebox{\hsize}{!}{\includegraphics{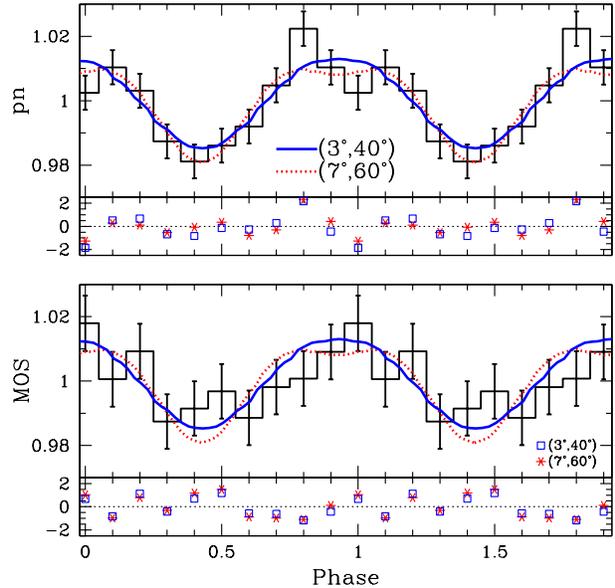}}
\caption{
Energy-integrated (0.15$-$1.2~keV) light curves of \rxjw.
Histograms are the pn (top) and MOS (bottom) observations (see Fig.~1 of TM07).
Solid and dotted lines are the models with ($3^\circ$,$40^\circ$) and
($7^\circ$,$60^\circ$), respectively, along with the fit deviations
[i.e., (data-model)/$\sigma$] in the corresponding lower panels.
Two rotation periods are shown for clarity.
}
\label{fig:fitpulse1}
\end{figure}

TM07 discovered 7~s pulsations in the 2006 October 24 observations
of \rxjw\ made by the EPIC-pn and MOS cameras on \xmm.
These observations cover the energy range 0.15$-$1.2~keV.
The pn light curve, with a pulse fraction of 1.6\%$\pm$0.2\%, is shown in
Figure~\ref{fig:fitpulse1} and found to have no significant
difference when divided into soft and hard energies;
the MOS light curve is also shown in Figure~\ref{fig:fitpulse1} and
found to be statistically consistent with no pulsations (TM07).

%-------------------------------------------
%-------------------------------------------

\begin{figure}
\resizebox{\hsize}{!}{\includegraphics{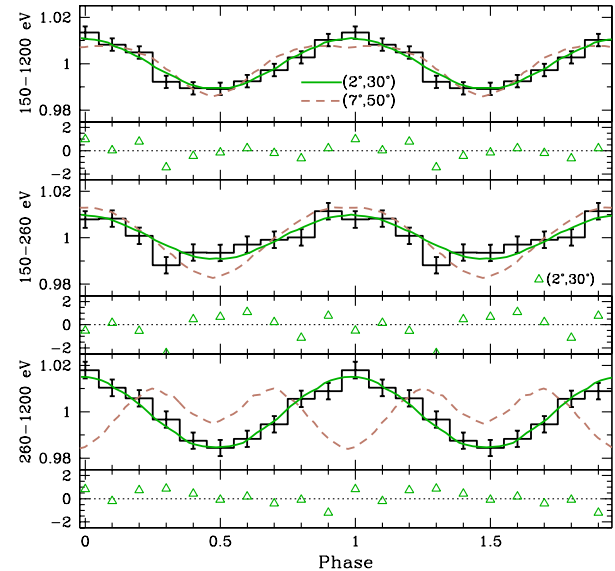}}
\caption{
Light curves of \rxjw\ for different energy ranges: 0.15$-$1.2~keV (top),
0.15$-$0.26~keV (middle), and 0.26$-$1.2~keV (bottom).
Histograms are the light curves obtained by summing together all the
\xmm\ observations (see Fig.~4 of TM07).
The solid lines are the model with ($2^\circ$,$30^\circ$), along with the fit
deviations [i.e., (data-model)/$\sigma$] in the corresponding lower panels,
while the dashed lines are the model with ($7^\circ$,$50^\circ$).
Two rotation periods are shown for clarity.
}
\label{fig:fitpulse2}
\end{figure}

As a result of this discovery, TM07 analyzed previous \xmm\ observations and
found light curves that are consistent between the different observations;
therefore these data were added together to obtain better statistics.
Furthermore, when divided into soft (0.15$-$0.26~keV) and hard
(0.26$-$1.2~keV) energy bands, the summed light curves
demonstrate some energy-dependence, i.e., pulse fractions of
$1.17\%\pm 0.08\%$ (0.15$-$1.2~keV),
$0.88\%\pm 0.11\%$ (0.15$-$0.26~keV), and
$1.5\%\pm 0.11\%$ (0.26$-$1.2~keV; see TM07).
The light curves are shown in Figure~\ref{fig:fitpulse2}.

%-------------------------------------------
%-------------------------------------------

Several characteristics of the observed light curves
suggest possible values of $\alpha$ and $\zeta$:
(1) the $\approx 1$\% amplitude of the pulsations,
(2) a single peak (or two peaks close in phase, as may be the case
for the pn light curve in Figure~\ref{fig:fitpulse1}) per rotation,
and (3) no significant energy-dependence for the single observation
(Figure~\ref{fig:fitpulse1}) and small energy-dependence for the combined
data (Figure~\ref{fig:fitpulse2}).
These characteristics imply the secondary cap is not a major contributor
to the brightness variation (i.e., not class III) and small values of
$\alpha$ or $\zeta$.  Large $|\alpha-\zeta|$ are also implied since
otherwise the pulsations would be strong with a narrow peak when the cap
becomes aligned with the direction to the observer
(a similar argument is made by Braje \& Romani~2002 from the non-detection
of \rxjw\ in the radio).

We compare the light curves generated from our model first to the single
(2006 October 24) \xmm\ observation and then to the summed data.
To avoid the computational tedious task of computing light
curves for a continuous distribution in the angles $\alpha$ and $\zeta$,
we select specific values of ($\alpha,\zeta$), compute the light curve, and
fit the model light curves to the observational data.
The phase and normalization of the model light curves are arbitrary
and are varied to minimize $\chisq$.
We restrict ourselves to integer values of ($\alpha,\zeta$) since the
uncertainties in our model (see H07 and Section~\ref{sec:surfdistrib})
and small variations between the data (c.f. Figures~\ref{fig:fitpulse1} and
\ref{fig:fitpulse2}) do not warrant more detailed fits.

We obtain good fits to the 2006 October 24 observation
with $\alpha$ and $\zeta$ of $\approx 2-9^\circ$ and $\sim 20-70^\circ$.
Table~\ref{tab:fitpulse} shows some of our fit results,
where the pn data is fit with a model light curve and then the MOS data
is fit using the same model light curve and phase alignment.
Figure~\ref{fig:fitpulse1} illustrates two sample fits.
These fits confine the allowed parameter space of ($\alpha,\zeta$).
Furthermore, the light curves from this observation do not show
significant differences when divided into soft and hard energies,
whereas the model light curves with $\zeta\gtrsim 50^\circ$ show
noticeably different hard-band pulsations.
Therefore, the energy-(in)dependence of the data argues for
$\zeta< 50^\circ$.

%-------------------------------------------
\begin{table}
\caption{Fits to 2006 October 24 Light Curves \label{tab:fitpulse}}
\begin{tabular}{c c c c}
\hline
$\alpha^{({\rm a})}$ & $\zeta^{({\rm a})}$
 & \multicolumn{2}{c}{$\chisq$/dof} \\
 (deg) & (deg) & pn$^{({\rm b})}$ & MOS$^{({\rm c})}$ \\
\hline
3 & 20 & 11.8/8 & 8.12/9 \\
3 & 30 & 11.4/8 & 8.36/9 \\
3 & 40 & 10.4/8 & 6.72/9 \\
9 & 50 & 9.94/8 & 9.92/9 \\
7 & 60 & 8.31/8 & 9.25/9 \\
2 & 70 & 11.2/8 & 7.48/9 \\
\hline
\end{tabular}
\begin{quote}
{\scshape Notes:}\\
(a) $\alpha$ and $\zeta$ are interchangeable. \\
(b) 10 phase bins $-$ 2 fit parameters (phase and normalization) = 8~dof \\
(c) The phase alignment from the pn fit is used.
\end{quote}
\end{table}
%-------------------------------------------

Finally, we fit the (energy-dependent) light curves obtained by adding
together the various \xmm\ observations.
Table~\ref{tab:fitpulse2} shows some of the results,
where the total (0.15$-$1.2~keV) data is fit with a model light curve
and then the soft (0.15$-$0.26~keV) and hard-band (0.26$-$1.2~keV) data
is fit using the same phase alignment.
The more stringent constraints on ($\alpha,\zeta$) are due to the
better statistics and energy-dependence of the combined data.
The model with ($\alpha,\zeta$)=($2^\circ,30^\circ$) is shown
in Figure~\ref{fig:fitpulse2}.

%-------------------------------------------
\begin{table}
\caption{Fits to Combined Light Curves \label{tab:fitpulse2}}
\begin{tabular}{c c c c c}
\hline
$\alpha^{({\rm a})}$ & $\zeta^{({\rm a})}$
 & \multicolumn{3}{c}{$\chisq$/dof} \\
 (deg) & (deg) & 0.15$-$1.2~keV$^{({\rm b})}$ & 0.15$-$0.26~keV$^{({\rm c})}$
 & 0.26$-$1.2~keV$^{({\rm c})}$ \\
\hline
2 & 20 & 4.88/8 & 12.2/9 & 9.42/9 \\
2 & 25 & 5.42/8 & 9.95/9 & 7.07/9 \\
2 & 30 & 4.43/8 & 10.4/9 & 3.88/9 \\
2 & 35 & 5.10/8 & 9.36/9 & 4.27/9 \\
2 & 40 & 8.34/8 & 9.85/9 & 6.25/9 \\
4 & 45 & 6.93/8 & 16.7/9 & 9.49/9 \\
\hline
\end{tabular}
\begin{quote}
{\scshape Notes:}\\
(a) $\alpha$ and $\zeta$ are interchangeable. \\
(b) 10 phase bins $-$ 2 fit parameters (phase and normalization) = 8~dof \\
(c) The phase alignment from the 0.15$-$1.2~keV fit is used.
\end{quote}
\end{table}
%-------------------------------------------

%-------------------------------------------
\begin{figure}
\resizebox{\hsize}{!}{\includegraphics{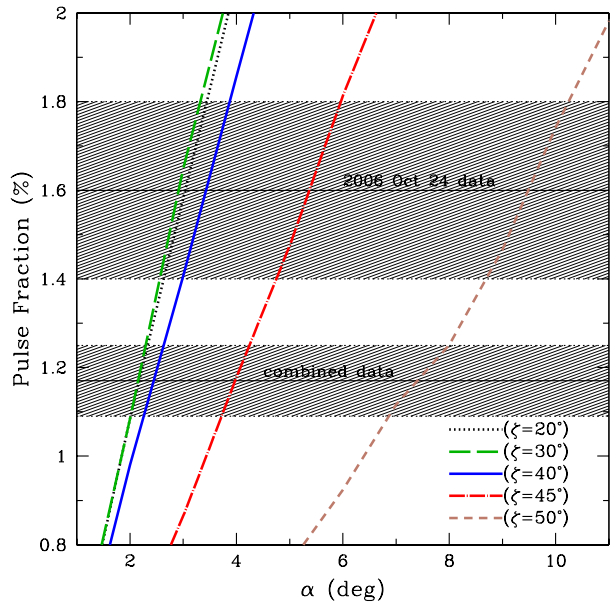}}
\caption{
Pulse fractions as a function of $\zeta$ (or $\alpha$) for different
$\alpha$ (or $\zeta$).
The horizontal lines indicate the pulse fractions from TM07
using data from the 2006 October 24 observation and the combined
observations; the shaded areas indicate the 1$\sigma$ uncertainty region.
}
\label{fig:pfang}
\end{figure}
%-------------------------------------------

As an illustration of our fit results, we show in Figure~\ref{fig:pfang} the
ranges allowed by the 1$\sigma$ uncertainty in the measured pulse fractions
and the pulse fractions of the model light curves as a function of $\alpha$
and $\zeta$.
Only model light curves whose pulse fractions lie in the shaded regions
can fit (within the uncertainties) the observed light curves.  However,
matching the pulse fraction is a necessary but not sufficient condition.
As discussed above, the pulse shape and energy-dependence also serve as
constraints.
For example, it appears from Figure~\ref{fig:pfang} that ($7^\circ,50^\circ$)
should fit the combined light curve.
But the shape of the model light curve for this case is too broad and
shows large differences when divided into soft and hard-energy bands
(see Figure~\ref{fig:fitpulse2}).

%%%%%%%%%%%%%%%%%%%%%%%%%%%%%%%%%%%%%%%%%%%%%%%%%%%%%%%%%%%%%%%%%%%
\section{Discussion} \label{sec:discussion}

In previous work (Ho \etal~2007), we successfully fit the multi-wavelength
spectrum of \rxjw\ with our model of radiation from a magnetic neutron
star containing a partially-ionized hydrogen atmosphere.
With the recent discovery of rotational modulation of the X-ray emission
(Tiengo \& Mereghetti~2007), we use the light curves predicted by our
model (which includes relatively simple surface distributions of the
magnetic field and temperature) to constrain the geometry of \rxjw.
In particular, we find angles of $< 6^\circ$ and
$\approx 20-45^\circ$: one is the angle $\alpha$ between the rotation and
magnetic axes and the other is the angle $\zeta$ between the rotation axis
and the direction to the observer.
Since the light curves with ($\alpha,\zeta$) are equivalent to those with
($\zeta,\alpha$), this implies that either the rotation and magnetic axes
are closely aligned or we are essentially seeing down the spin axis of the
neutron star (see also Braje \& Romani~2002).
We also note the following:
(1) The model used here and described in Ho \etal~(2007) assumes the high
X-ray energy emission arises from a condensed iron surface below the
hydrogen atmosphere.  The emission properties of the condensed surface
have been studied by Brinkmann~(1980); Turolla, Zane, \& Drake~(2004);
P\'{e}rez-Azor\'{i}n, Miralles, \& Pons~(2005); van Adelsberg \etal~(2005).
(2) Though it would present an observational challenge, our model predicts
optical pulsations with a singly-peaked sinusoid and pulse fractions of
$\approx 0.1-0.5\%$.
(3) Other members of the neutron star population \rxjw\ is thought to
belong to show varying degrees of pulsations, some with a single peak
(per rotation) and others with double peaks.  Analyses of these sources
will be presented in future work.

%%%%%%%%%%%%%%%%%%%%%%%%%%%%%%%%%%%%%%%%%%%%%%%%%%%%%%%%%%%%%%%%%%%
\subsection{Changes in Surface Distributions of $B$ and $\Teff$}
\label{sec:surfdistrib}

We point out an uncertainty in our model that is difficult to quantify
but will affect the light curves and thus the quality of the fits
and the angles ($\alpha,\zeta$) inferred.
We assume dipolar-like distributions of the magnetic field and temperature
over the surface of the neutron star (see Table~\ref{tab:nssurf}).
If the size of each magnetic colatitudinal region is varied, the strength
of the pulsations will change.  For the geometries that best fit the
observations, a $\pm 5^\circ$ change in the size of the
the region described by $B=4\times 10^{12}$~G, $\ThetaB=60^\circ$, and
$\Teff=5\times 10^5$~K can result in a $\lesssim 0.002$ change in the
pulse fraction, which is on the order of the observational errors.
In addition, if the distributions are described by an offset dipole or if
quadrupolar components are added on top of the dipole, then the shape and
strength of the pulsations will, in general, be different.

%%%%%%%%%%%%%%%%%%%%%%%%%%%%%%%%%%%%%%%%%%%%%%%%%%%%%%%%%%%%%%%%%%%
\subsection{\rxjw\ as a Normal Radio Pulsar?} \label{sec:pulsar}

One set of our solutions ($\alpha < 6^\circ$) indicates a very close
alignment between the rotation and magnetic axes.
Observational evidence from studies of radio pulsars may also indicate the
alignment of the rotation and magnetic axes on a timescale of $\sim 10^7$~y
(Lyne \& Manchester~1988; Tauris \& Manchester~1998).
For example, Tauris \& Manchester~(1998; see also Zhang, Jiang, \& Mei~2003,
and references therein) find the observed distribution of $\alpha$ peaks
around $30-50^\circ$; after correcting for beaming, they find $\alpha$
decreases with time, with an average $\alpha\approx 40^\circ$ for
pulsars younger than $10^{6.5}$~y.
Thus both our solutions ($\alpha<6^\circ$ and $\alpha\approx 20-45^\circ$)
fit within this scenario, given the $\sim 5\times 10^5$~y age of \rxjw\
(Kaplan, van Kerkwijk, \& Anderson~2002; Walter \& Lattimer~2002),

In spite of its non-detection at radio frequencies (Brazier \& Johnston~1999),
suppose \rxjw\ is a normal radio pulsar.
The size of the radio beam $\rho$ is estimated to be
$\rho\approx 5-6^\circ(P/\mbox{1 s})^{-1/2}=2^\circ$
(see, e.g., Rankin~1993; Gould~1994).
Our results allow the radio beam to miss our line-of-sight, with the
($\alpha\approx 20-45^\circ,\zeta<6^\circ$)-solution having a easier
chance for this to occur.
In addition, if \rxjw\ is losing rotational energy by magnetic dipole
radiation, the spindown rate is
$\dot{P}=10^{-15}\mbox{ s s$^{-1}$}(B/\mbox{10$^{12}$ G})^2
(P/\mbox{1 s})^{-1}=5\times 10^{-15}\mbox{ s s$^{-1}$}$,
which is well below the $\dot{P} < 1.9\times  10^{-12}\mbox{ s s$^{-1}$}$
obtained by Tiengo \& Mereghetti~(2007).
This spindown rate (and $\sim 5\times 10^5$~y age) implies that \rxjw\
must have been born at roughly
its current spin period if no significant change in $\dot{P}$ (or $B$)
occurred over its lifetime.
Also, an H$\alpha$ nebula was found around \rxjw\ and has an
estimated luminosity of $\dot{E}_{\rm bow}=8\times 10^{32}$~ergs~s$^{-1}$
(van Kerkwijk \& Kulkarni~2001b);
the dipole spindown luminosity of \rxjw\ is insufficient in powering
this relativistic wind.
However, it is important to note in the above considerations that,
given its spin period and magnetic field, \rxjw\ falls below the death
line for radio pulsars;
the death line being the region in $P-B$-space (or $P-\dot{P}$)
where pulsars no longer emit in the radio (see, e.g., Sturrock~1971;
Chen \& Ruderman~1993; Harding, Muslimov, \& Zhang~2002).

%%%%%%%%%%%%%%%%%%%%%%%%%%%%%%%%%%%%%%%%%%%%%%%%%%%%%%%%%%%%%%%%%%%
\subsection{X-ray Polarization} \label{sec:xraypol}

In the presence of magnetic fields typical of isolated neutron
stars such as \rxjw\ ($B\ga 10^{12}$~G), radiation propagates in two
polarization modes:
the ordinary mode is polarized parallel to the $\veck$-$\vecB$ plane,
while the extraordinary mode is polarized perpendicular to the
$\veck$-$\vecB$ plane (see, e.g., M\'{e}sz\'{a}ros~1992).  Because the
extraordinary mode opacity is greatly suppressed compared to the ordinary
mode opacity [by a factor $\sim (E/E_B)^2$; Lodenquai \etal~1974;
M\'{e}sz\'{a}ros~1992], the extraordinary mode photons
escape from deeper, hotter layers of the atmosphere than the ordinary mode
photons, and the emergent radiation is linearly polarized to a high degree
(as high as 100\%; Gnedin \& Sunyaev~1974; M\'{e}sz\'{a}ros \etal~1988;
Pavlov \& Zavlin~2000).
Measurements of X-ray polarization, particularly when phase-resolved and
measured in different energy bands, could provide unique constraints on
the magnetic field strength and geometry and the compactness of the neutron
star (M\'{e}sz\'{a}ros \etal~1988; Pavlov \& Zavlin~2000;
Heyl, Shaviv, \& Lloyd~2003; Lai \& Ho~2003).
As noted previously, light curves of the total flux with
($\alpha,\zeta$) are equivalent to those with ($\zeta,\alpha$).  Thus the
results presented in this work
could be satisfied equally with an interchange of these two angles.
However, the light curves of the polarized light for the two sets of angles
are distinctly different (Pavlov \& Zavlin~2000), with the ordinary mode
spectra showing greater phase variability since the outer atmosphere layers
are more sensitive to magnetic field orientation (Ventura \etal~1993).
For our case, the
polarization is essentially 100\% in one polarization, and the position
angle (in the plane of the sky) of the linear polarization would undergo
small changes as the star rotates for the case with ($2^\circ,30^\circ$),
while the position angle would rotate $360^\circ$ for the case with
($30^\circ,2^\circ$).
Propagation effects, such as those discussed in Heyl \& Shaviv~(2000, 2002);
Heyl \etal~(2003); Lai \& Ho~(2003), would need to be taken into account,
but sources like \rxjw\ could be studied by possible future X-ray
polarization instruments, such as {\it Constellation-X}, {\it XEUS}, and
the {\it Extreme Physics Explorer}
(Bellazzini \etal~2006; Elvis~2006; Jahoda \etal~2007).

\section*{acknowledgements}

WH is grateful to Andrea Tiengo for providing the \xmm\ light curves
and David Kaplan, Kaya Mori, Patrick Slane, and Marten van Kerkwijk for
useful comments on an early version of the paper.
WH thanks the anonymous referee for helping to improve the clarity
of the paper.
WH appreciates the use of the computer facilities at the Kavli
Institute for Particle Astrophysics and Cosmology.

%%%%%%%%%%%%%%%%%%%%%%%%%%%%%%%%%%%%%%%%%%%%%%%%%%%

\label{lastpage}

\end{document}